\documentclass[twoside,11pt]{article}
\usepackage{jmlr2e_draft}

\usepackage{amsmath, amssymb} 
\usepackage{stmaryrd}
\usepackage{epsfig,calc, graphicx}
\usepackage{wasysym}
\usepackage{enumitem}
\usepackage{scalerel}

\usepackage{bbm}
\usepackage{comment}
\usepackage{hyperref}

\usepackage{array}
\usepackage{booktabs}

\newcommand\absip[2]{\left\lvert\langle #1, #2\rangle\right\rvert}

\newcommand\dico{A}
\newcommand\pdico{\Psi}

\newcommand\atom{{\phi}}
\newcommand\patom{{\psi}}

\newcommand{\transp}{^{\star}}

\newcommand\diag{\operatorname{diag}}
\newcommand\signop{\operatorname{sign}}

\newcommand{\N}{{\mathbb{N}}}
\newcommand{\R}{{\mathbb{R}}}

\newcommand{\C}{{\mathbb{C}}}
\newcommand{\E}{{\mathbb{E}}}

\renewcommand{\P}{{\mathbb{P}}}

\newcommand{\sampleprob}{\pi}
\newcommand{\poissonprob}{p}

\newcommand{\weightsp}{D_{\scaleto{ \sqrt{p}\mathstrut}{6pt}}}

\newcommand\proj{D_I}

\begin{document}




\author{\name Simon Ruetz \email simon.ruetz@yahoo.de\\
	\addr  
	University of Innsbruck\\
	Technikerstra\ss e 13\\
	6020 Innsbruck, Austria}

\title{Adapted variable density subsampling for compressed sensing}

\maketitle

\begin{abstract}
Recent results in compressed sensing showed that the optimal subsampling strategy should take the sparsity pattern of the signal at hand into account. This oracle-like knowledge, even though desirable, nevertheless remains elusive in most practical applications. We try to close this gap by showing how the sparsity patterns can instead be characterised via a probability distribution on the supports of the sparse signals allowing us to again derive optimal subsampling strategies. This probability distribution can be easily estimated from signals of the same signal class, achieving state of the art performance in numerical experiments. Our approach also extends to structured acquisition, where instead of isolated measurements, blocks of measurements are taken.
\end{abstract}

\section{Compressed Sensing}
\noindent

Let $x \in \C^{K}$ be some signal and $A \in \C^{m \times K}$ be some matrix. Compressed sensing (CS) consists of reconstructing the signal $x$ from measurements $y = Ax$. Usually it is assumed that the signal $x$ is $S$-sparse, meaning that only $S\ll K$ elements of $x$ are non-zero. Thus one tries to solve the following optimisation problem
\begin{equation}\label{BP}
\hat{x} = \arg \min \| x \|_1  \quad \text{s.t.} \quad y = A x.
\end{equation}
Starting with the seminal works~\cite{candes06,doelte06}, compressed sensing theory tries to find sufficient conditions for the above minimisation problem to recover the the sparse signal. Early results suggested that if each entry of the matrix $A$ is sampled i.i.d. from a Gaussian distribution and $m \gtrsim S \log(K)$, then the above minimisation does yield the correct solution with high probability.\\
These results were soon extended to a random subsampling setting, where the sensing matrix $A$ is constructed by sampling rows $a_k$ from a unitary matrix $A_0 \in \C^{K \times K}$ uniformly at random~\cite{candes11, ra10}. In this setting, a typical sufficient condition for the above minimisation problem to recover the sparse signal with probability at least $1 - \varepsilon$ reads as
\begin{equation}
    m \gtrsim S K \max_{1 \leq k \leq K}\|a_k\|^2_{\infty} \log(K/\varepsilon).
\end{equation}
If $A_0$ is the discrete Fourier matrix --- for which $\max_{1 \leq k \leq K}\|a_k\|^2_{\infty} = \frac{1}{K}$ --- this leads to theoretical results comparable to the Gaussian setting. Nevertheless this still falls short of explaining the remarkable success of CS in most applications where $K \max_{1 \leq k \leq K}\|a_{k}\|^2_{\infty}$ is usually quite large. \\
To solve this problem, variable density subsampling was introduced~\cite{ra10, weiss13, Vander11, weiss13_2,krwa12, bibocl14}. There the sensing matrix $A \in \C^{m \times K}$ is constructed by sampling the rows of $A_0$ via a (possibly non-uniform) probability distribution. Concretely, the sensing matrix $A$ is defined to be
\[
A: =  \frac{1}{\sqrt{m}}\left(\frac{1}{\sqrt{\pi_{j_{\ell}}}} a_{j_{\ell}} \right)_{1 \leq \ell \leq m}, 
\]
where $m$ is the number of measurements we are allowed to take and $j_{\ell}$ for $1 \leq \ell \leq m$ are i.i.d random variables such that $\P( j_{\ell} = k) = \pi_k$. Note that the subsampling strategy is determined by the probabilities $\pi_k$ for $1 \leq k \leq K$. A typical choice in this setting is $\pi_k := \frac{\|a_k\|^2_{\infty}}{\sum_k \|a_k\|^2_{\infty}}$ leading to the sufficient condition
\begin{equation*}
    m \gtrsim S \sum_k \|a_k\|^2_{\infty} \log(K/\varepsilon).
\end{equation*}
\\
This is nevertheless still not enough to completely bridge the gap between theory and application. Recent results go further by arguing that the optimal subsampling strategy should not only depend on the sensing and sparsity matrices, but also on the structure of the sparse signals~\cite{adhaporo17, pierre15,adcock20}. The so called flip test proposed in~\cite{adhaporo17} is a prime example of this fact. The assumption of knowledge of the structure of the sparse signals was also shown to be especially important in the case of blocks of measurements~\cite{pierre15, adcock17, adcock20}. The drawback of all of these results is that they rely on the exact knowledge of the locations of the non-zero coefficients of the sparse signal, which may not be available in practice. 

\section{Contribution}
In this paper we generalise these results to show how the subsampling strategy depends on the \textbf{distribution} of sparse supports together with the structure of the sensing/sparsity matrix. 
We are able to do this by assuming that the sparse supports follow a (possibly) non-uniform distribution, thereby generalising the aforementioned results. In practice, if one has access to a number of \textit{similar} signals to $x$, a guess of the underlying distribution of sparse supports of $x$ can be made and the optimal subsampling pattern be thus estimated. We also extend our results to the setting of structured acquisition, where instead of isolated measurements, blocks of measurements are taken. In Section~\ref{sec:notations} we introduce the relevant notation, Section~\ref{sec:main} states the main result, Section~\ref{sec:application} shows how to apply this result in practice and Section~\ref{sec:special} applies our theory to some special cases to compare it to existing results. The proof of our main result is deferred to appendix~\ref{sec:proof}.

\section{Notation and setting}\label{sec:notations}
A quick note on the notation used throughout this text. For an integer $K$, we write $\mathbb{K} := \{1 ,\cdots, K\}$. The vectors $(e_i)_{1\leq i \leq K}$ denote the vectors of the canonical basis of $\R^K$. For a matrix $A \in \C^{d \times K}$, we denote by $A_{:,k}$ (resp. $A_{k,:}$) the $k$-th column (resp. row) of $A$ and by $A_{J,L}$ the submatrix with rows indexed by set $J\subseteq \{1, \cdots , d\} $ and columns indexed by set $L \subseteq \mathbb{K}$. If we talk about certain columns of a matrix $A$, we often drop the second index, i.e. instead of $A_{:,k}$ we will write $A_k$ and instead of $A_{:,J}$, we will write $A_J$. By $A\transp$ we denote the conjugate transpose of the matrix $A$ and by $A_k \transp \in \R^{1 \times d}$, the conjugate transpose of the $k$-th column of $A$.\\
For $1 \leq p,q,r \leq \infty$ we set $$\| A \|_{p,q} := \max_{\| x\|_q = 1} \| Ax \|_p. $$
So for $B \in \C^{K\times m}$ we get $\| A B \|_{p,q} \leq \|A \|_{q,r} \|B\|_{r,p}$ and $\|Ax \|_{q} \leq \|A\|_{q,p} \|x \|_{p}$. Frequently encountered quantities are
\[
\| A \|_{\infty, 2} = \max_{k \in \{1, \dots , d \}}\|A_{k,:} \|_2 \quad \text{and} \quad \| A \|_{2, 1} = \max_{k \in \{1, \dots , K \}}\|A_k \|_2,
\]
which denote the maximum $\ell_2$-norm of a row and the maximum $\ell_2$-norm of a column of $A$ respectively. Note that $\| A \|_{\infty,2} = \|A\transp \|_{2,1}$. Further note that $\| A \|_{\infty,1}$ is the maximum absolute entry of the matrix $A$. For ease of notation we sometimes write $\| A \| = \| A \|_{2,2}$ for the operator norm which corresponds to the largest singular value of $A$. For a vector $v \in \R^{d}$, we denote by $\underline{v} := \|v\|_{\min} := \min_{i} \vert v_{i}\vert $ the smallest absolute entry of $v$ and $\|v\|_{\max} := \|v \|_{\infty}$ the biggest absolute entry of $v$. We write $x \lesssim y$ if there exists a constant $c>0$, such that $x \leq c y$. We write $\operatorname{vec}:\C^{d \times d} \mapsto \C^{d^2}$ for the vectorisaton operation that transforms a complex matrix into a complex vector by stacking the columns on top of each other and by $\operatorname{vec}^{-1}$ its inverse. 
Further, for any vector $v$ we denote by $D_v$ resp. $\diag(v)$ the diagonal matrix with $v$ on the diagonal.\\
As was noted in the introduction we want the supports of our signals to follow a non-uniform distribution. We are going to use the following probability measure on $\mathcal{P}(\mathbb{K})$ that allow us to model non-uniform distributions for our supports while still ensuring that they are $S$-sparse.
\begin{definition}[Rejective sampling - Conditional Bernoulli model]
Let $0 \leq \poissonprob_j \leq 1$ be such that $\sum_{j =1}^K \poissonprob_j = S$. We say our supports follow the rejective sampling model, if each support $I \subseteq \mathbb{K}$ is chosen with probability
\begin{equation}\label{cond_dist}
    \P(I) := \begin{cases}
               c \prod_{i \in I}\poissonprob_i\prod_{j \notin I}(1-\poissonprob_j) \quad& \mbox{if} \quad \lvert I \rvert=S\\
               0 \quad& \mbox{else}
            \end{cases},
\end{equation}
where $c$ is a constant to ensure that $\P$ is a probability measure. We define the $D_{\poissonprob} := \diag((\poissonprob_k)_k)$ as the square diagonal matrix with the weight vector $\poissonprob$ on its diagonal. We call $W \in \R^{\sqrt{K}\times \sqrt{K}}$ the weight matrix, if $\operatorname{vec}(W) = \poissonprob$\footnote{Here we implicitly assume that $\sqrt{K}$ is an integer.}.
\end{definition}
One important thing to note here is that in general $p_i \neq \E[i \in I]$ (except for the case $p_i = \frac{S}{K}$). Luckily, especially for large $K$, we have $p_i \approx \E[i \in I]$~\cite{hajek1964,ruetz2022nonasymptotic}, which allows one to estimate the underlying probability vector $p$ by approximating them by the probability of occurrence $\E[i \in I]$ which in turn can be estimated from a dataset.\\
Let $\proj$ be the square diagonal \textit{selector matrix} satisfying $(\proj)_{{\ell},{\ell}} = 1  \Leftrightarrow {\ell} \in I $.

\noindent This lets us define the following model for our signals. We specify two models, one for the complex and one for the real case.
\begin{definition}[Signal model]\label{signal_model}
We model our signals as
\begin{align}
    x = \sum_{i \in I} e_{i} x_{i} \sigma_{i} ,
\end{align}
where $x_i \in \R$ (or $\C$) and $I = \{i_1, \dots i_S \}$ is the random support following the rejective sampling model with weight vector $\poissonprob$ such that $\sum_{i = 1}^K \poissonprob_i = S$ and denote by $D_{\poissonprob}$ the corresponding diagonal matrix. Further we assume that $\sigma_i$ forms a Rademacher sequence.
\end{definition}

\section{Main result}\label{sec:main}
Assume we are given a unitary matrix $A_0  \in \C^{K \times K}$ representing the set of possible linear measurements $(A_0 \transp)_i =: a_i\transp$. We partition the set $\mathbb{K}$ into $M$ blocks $\mathcal{I}_k$ such that $\uplus_k \mathcal{I}_k = \mathbb{K}$ and set
\[
B_k := (a_i)_{i \in \mathcal{I}_{k}} \in \C^{\lvert \mathcal{I}_k \rvert \times K}
\]
The sensing matrix $A$ is then defined as
\[
A: =  \frac{1}{\sqrt{m}}\left(\frac{1}{\sqrt{\sampleprob_{j_{\ell}}}} B_{j_{\ell}} \right)_{1 \leq \ell \leq m}, 
\]
where $m\leq M$ is the number of blocks we want to measure and $j_{\ell}$ for $1 \leq \ell \leq m$ are i.i.d random variables such that $\P( j_{\ell} = k) = \sampleprob_k$. So the $\sampleprob_k$ define the probability with which each block of measurements is selected. In line with existing compressed sensing literature we call 
\begin{align}
    \max_k \|a_k\|^2_{\infty},
\end{align}
the coherence of the matrix $A_0$. With these definitions we are finally able to state our main result.
\begin{theorem}\label{them:1}
Assume $m$ measurements $B_k$ are sampled according to probabilities $\sampleprob_k$ and that the signals follow the model in~\ref{signal_model}, where the support $I \subseteq \mathbb{K}$ is chosen according to the rejective sampling model with probabilities $\poissonprob_1, \cdots, \poissonprob_K$ such that $\sum_{k =1}^K \poissonprob_k = S$ and $0 < \poissonprob_k \leq 1$. If
\begin{align}
    m &\gtrsim \max_{k}  \frac{\| B_{k}\transp B_{k} \|_{\infty,1} }{\sampleprob_{k}} \log^3(K/\varepsilon), \nonumber\\
    m &\gtrsim \max_{k} \frac{\| B_{k} D_{\poissonprob} B_{k}\transp \|_{2,2}}{\sampleprob_{k}}   \log^2( K/\varepsilon) \label{bound2},
\end{align}
then \eqref{BP} recovers the sparse signal with probability $1- \varepsilon$.
\end{theorem}
The exact statement --- including constants --- can be found in Section~\ref{sec:proof}. The restriction $\poissonprob > 0$ is no real constraint, as in the case of $\poissonprob_i = 0$ for some $i$, a careful analysis of the proof shows that one can then set the $i$-th column of $A$ to zero since this index is never part of the random supports $I$ anyway.
\begin{remark}
This result also extends to signals $x$ that are sparse in some unitary basis $\pdico$ by a change of variable. If we denote by $\Phi\transp$ our original sensing matrix and let $x = \pdico z$ for some sparse vector $z$, then we can again apply the above result with the new sensing matrix $A_0 = \Phi \transp \pdico$ and sparse signal $z$. In this case, the coherence $\|a_k\|_{\infty}$ is similar to a cross-coherence by noting that $\|a_k\|_{\infty} = \max_{i,j}\absip{\atom_i}{\patom_j}$.
\end{remark}
\begin{remark}
Even thought the above result is stated in terms of the product probability measure of measurements and signals, careful analysis of the proof shows that in fact it can also be understood in a sequential manner. When sampling the measurement matrix according to probabilities $\pi_k$ (and the conditions of the theorem are satisfied), then with high probability one will get a measurement matrix that works for most signals. This is what one hopes for since the aim is usually to construct a measurement matrix that works well for multiple signals.
\end{remark}
The above result shows that the optimal sampling strategy $\sampleprob$ should depend both on the distribution $\poissonprob$ of sparse supports via the diagonal matrix $D_{\poissonprob}$ and on the structure of the blocks~$B_k$. 
One way to optimise the bounds is by setting
\begin{equation}
    \sampleprob_k := \frac{\max\left\{\|B_k D_{\poissonprob} B_k \transp \|_{2,2}, \|B_k \transp B_k \|_{\infty,1} \right\}}{L}, \label{formula_blocks}
\end{equation}
where $L$ is a normalising constant ensuring $\sum_k \sampleprob_k = 1$. By plugging this bound into the above theorem we get the sufficient condition
\begin{align}
    m \gtrsim \left(  \sum_k \|B_k D_{\poissonprob} B_k \transp \|_{2,2} + \sum_k \|B_k \transp B_k \|_{\infty,1} \right) \log^3(K/\varepsilon).\label{block_bound}
\end{align}
In Section~\ref{sec:special} we will look at special cases of blocks of measurements, where this bound on $m$ can further be simplified.
For isolated measurements, i.e. the case where $B_k = a_k$, the above yields the following result.
\begin{corollary}\label{lem}
Assume that the signals follow the model in~\ref{signal_model}, where the support $I \subseteq \mathbb{K}$ is chosen according to the rejective sampling model with probabilities $\poissonprob_1, \cdots, \poissonprob_K$ such that $\sum_{k =1}^K \poissonprob_k = S$ and $0 < \poissonprob_k \leq 1$. If the measurements $a_k$ are sampled according to
\begin{align}
    \sampleprob_k = \frac{\max\{ a_k D_{\poissonprob} a_k \transp, \|a_k\|^2_{\infty}\}}{L}\label{formula},
\end{align}
where $L$ is a normalising constant ensuring $\sum_k \sampleprob_k =1$, and if
\begin{align}
    m \gtrsim \left( S + \sum_k\| a_k \|^2_{\infty}\right) \log^3(K/\varepsilon),
\end{align}
then \eqref{BP} recovers the sparse signal with probability $1- \varepsilon$.
\end{corollary}
\begin{proof}
First note that $\|B_k D_{\poissonprob} B_k \transp \|_{2,2} =  a_k D_{\poissonprob} a_k \transp$ and thus 
\begin{equation*}
    \sum_k a_k D_{\poissonprob} a_k \transp  = \operatorname{tr}( A_0 D_{\poissonprob} A_0 \transp) = \operatorname{tr}(D_{\poissonprob}) = S.    
\end{equation*}
Further 
\begin{equation*}
    \|B_k \transp B_k\|_{\infty,1} = \| a_k \transp a_k \|_{\infty,1} \leq \max_{i,j} \lvert a_{k,i} a_{k,j} \rvert \leq \max_i \lvert a_{k,i} \rvert^2 = \|a_k \|^2_{\infty},
\end{equation*}
leading to $L \leq S + \sum_k\| a_k \|^2_{\infty}$. Plugging these $\sampleprob_k$ into Theorem~\ref{them:1} yields the result.
\end{proof}
This result is an improvement upon standard results for general (unknown) supports $I$, which read as $m \gtrsim S \sum_k\| a_k \|^2_{\infty} \log(K)$~\cite{candes11, Vander11, krwa12,weiss13_2}. This is to be expected since we assume that information about the supports and their distribution is available. On the other hand, the additional log factors are the price we pay for our random signal approach. A comparison to existing results that assume knowledge about the structure of sparsity will be conducted in Section~\ref{sec:special}.\\
Further, Corollary~\ref{lem} shows how, for a given weight vector $\poissonprob$, this lower bound is attained via the formula in~\eqref{formula}. This is an easy-to-use recipe yielding state of the art results in a number of experiments (see also Sections~\ref{sec:special}). We now show that this indeed outperforms standard subsampling procedures in numerical experiments. 

\section{Numerical experiments}\label{sec:application}
\begin{figure}[ht]
    \centering
    \includegraphics[width=0.66\linewidth]{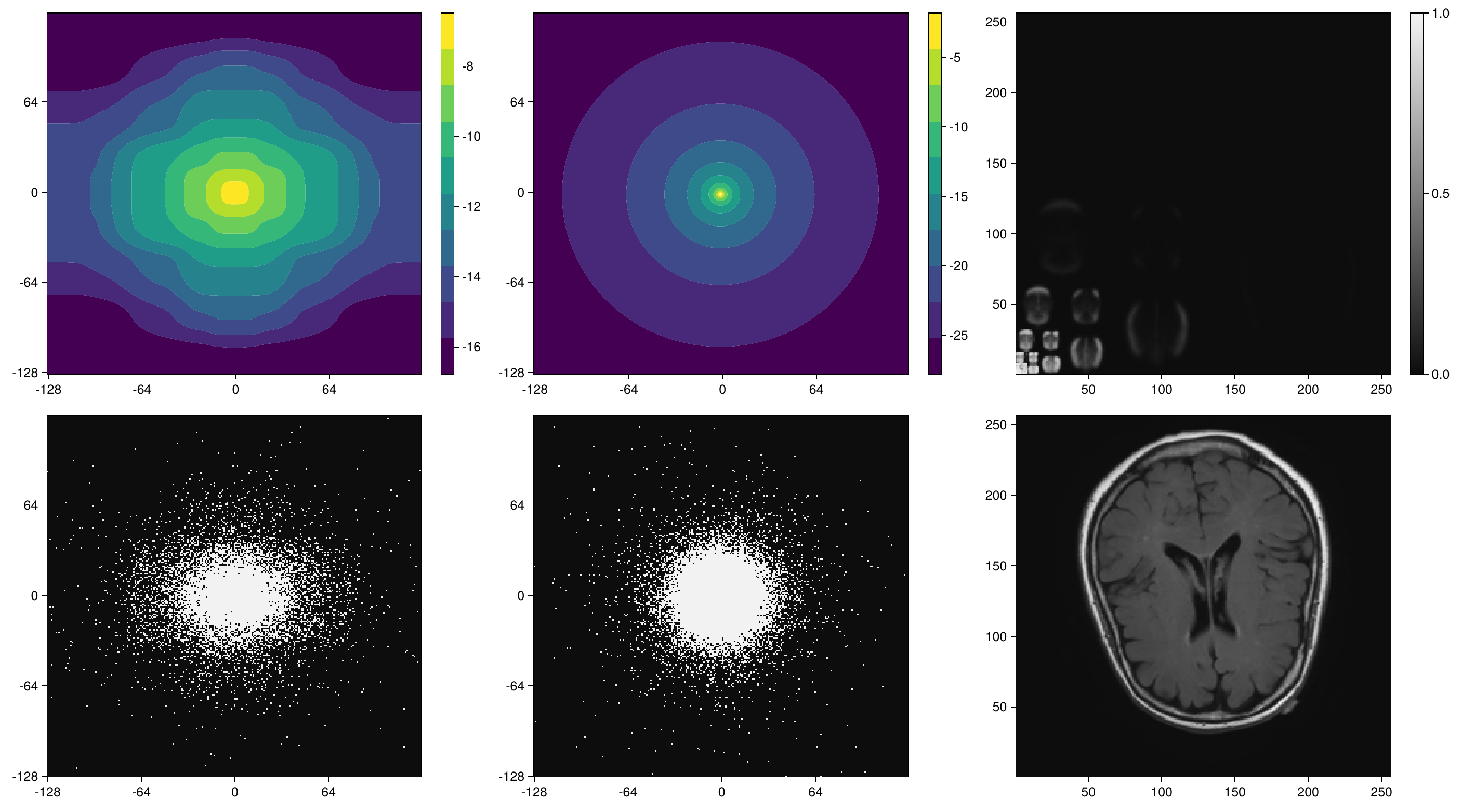}
    \caption{Adapted variable density sampling scheme (left column) vs polynomial decay (middle column). Estimated Matrix $W$ of inclusion probabilities for the sparse support distribution in the DB4 wavelet basis (top right) and test image (bottom right). The resulting PSNR values are: Adapted - 32.8 and Polynomial - 32.0.}
    \label{fig:wavelet}
\end{figure}

\begin{figure}[ht]
    \centering
    \includegraphics[width=0.66\linewidth]{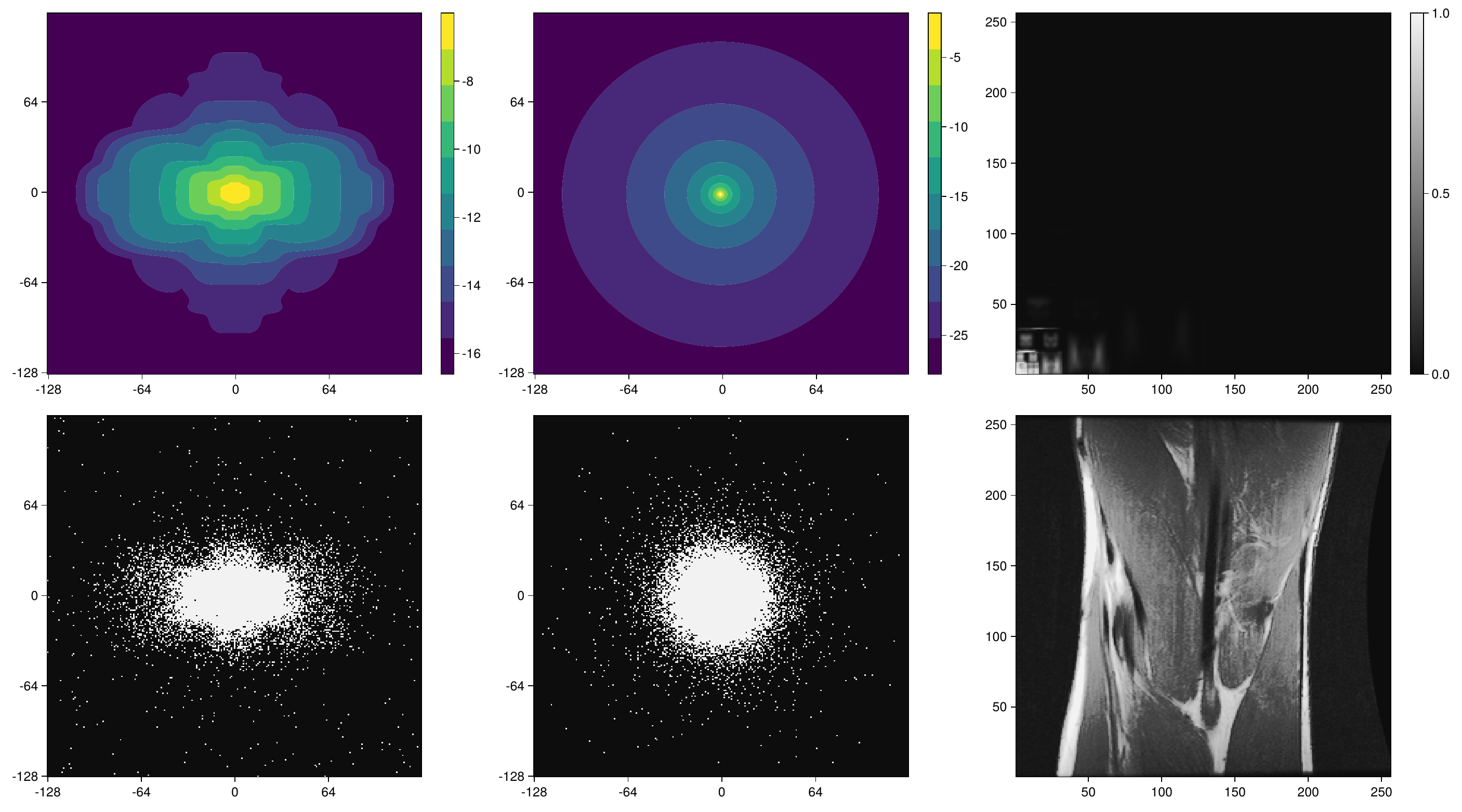}
    \caption{Adapted variable density sampling scheme (left column) vs polynomial decay (middle column). Estimated Matrix $W$ of inclusion probabilities for the sparse support distribution in the DB4 wavelet basis (top right) and test image (bottom right). The resulting PSNR values are: Adapted - 27.9 and Polynomial - 26.8.}
    \label{fig:knee}
\end{figure}

\begin{figure}[ht]
    \centering
    \includegraphics[width=0.66\linewidth]{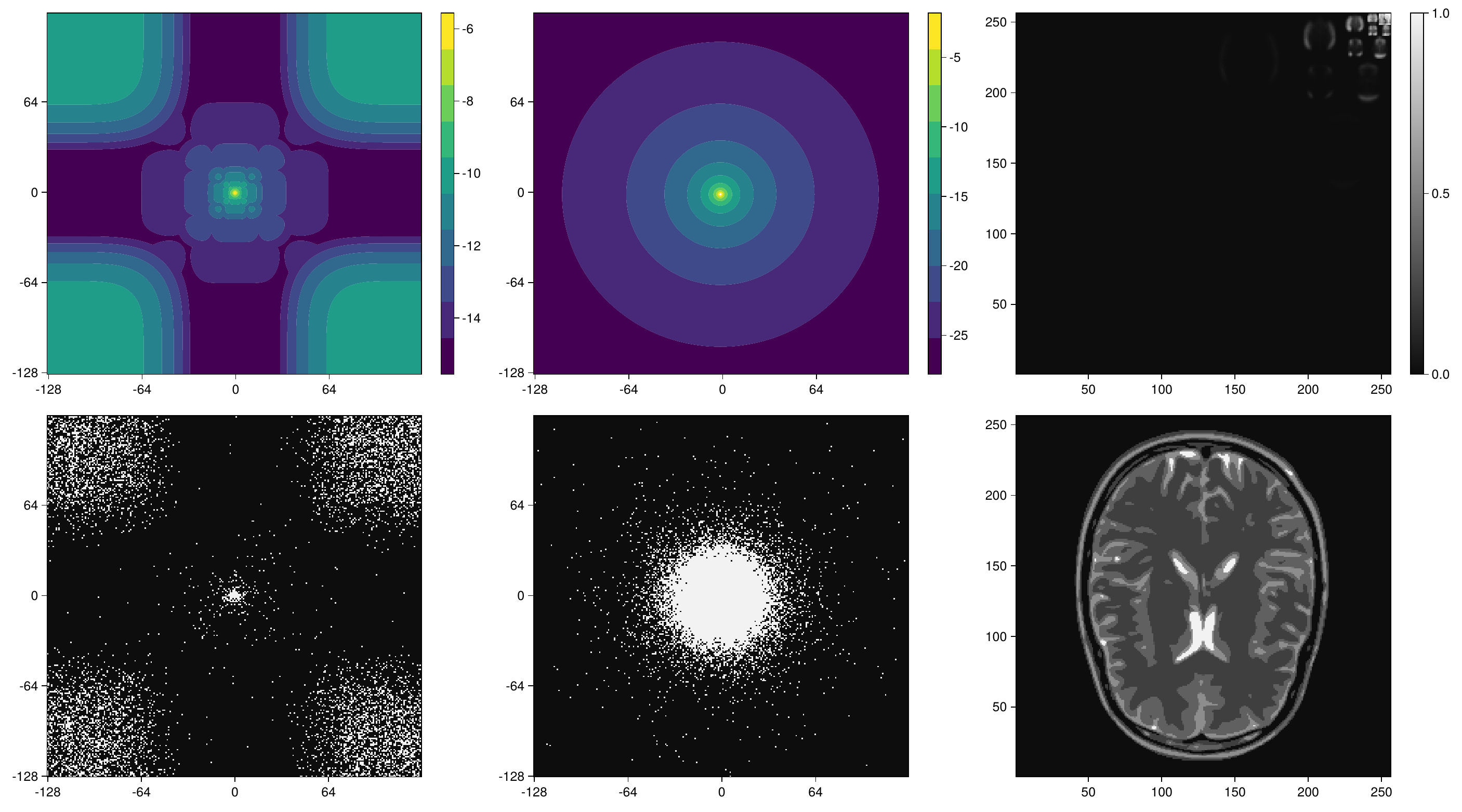}
    \caption{Adapted variable density sampling scheme (left column) vs polynomial decay (middle column). Estimated Matrix $W$ of inclusion probabilities for the sparse support distribution in the DB4 wavelet basis (top right) and test image (bottom right). The resulting PSNR values are: Adapted - 26.4 and Polynomial - 14.0.}
    \label{fig:flip}
\end{figure}

Now we conduct a few experiments to compare the performance of our subsampling scheme to heuristically inspired subsampling schemes used in practice. For our first experiment (Figure~\ref{fig:wavelet}) we assume a standard compressed sensing setup with isolated 2D Fourier measurements and a 2D DB4 wavelet matrix as sparsifying basis.\\
We assume to be given a training set of images from which we generate the sparse distribution model by transforming them into a wavelet basis before applying a threshold. The relative frequency with which each coefficient appears in these sparse supports is our proxy for the inclusion probabilities $\poissonprob$, since for large sample sizes they are approximately equal to the expectation of an atom being in the support~\cite{hajek1964,ruetz2022nonasymptotic}. This one-to-one correspondence is related to the close relationship between the rejective sampling model and the Bernoulli sampling model with weights $\poissonprob$~\cite{hajek1964,ruetz2022nonasymptotic,rusc21}.\\
We further assume to be given a reference image (bottom right) which we have to reconstruct. We will compare the performance of our subsampling strategy in the isolated measurement case against a state-of-the-art variable density subsampling scheme with polynomial decay, where we pick a frequency $(k_1,k_2)$ in the 2D k-space with probability $\frac{1}{(k_1^2 + k_2^2)^{2.5}}$. To ensure meaningful results, each experiment is averaged over 10 runs. All sampling distributions will be plotted in log-scale. \\
To approximate the distribution of the sparse supports, we use a dataset of around $4.000$ real brain images~\cite{dataset_brains} onto which we apply the $2$D DB4 wavelet transform followed by a thresholding operation with a threshold of around $0.006$, yielding the weight matrix $W$ (top right). Plugging these weights into Formula~\eqref{formula} and normalising the resulting density to $1$, we get the adapted subsampling distribution $\sampleprob$ (top left). We compare this strategy to the above mentioned polynomial decaying density (top middle) by sampling $10\%$ of frequencies in the k-space (bottom left and middle). Finally, an application of the Nesta algorithm to solve~\eqref{BP} for both sets of measurements yields the results in the figure. As can be seen, the adapted subsampling strategy is able to slightly outperform the quadratically decaying subsampling strategy --- resulting in a PSNR value of 23.8 compared to 32.0.\\
\noindent To show that our new subsampling strategy does indeed adapt to the underlying distribution of sparse supports, we repeat the above experiment (Figure~\ref{fig:knee}) but this time use a different dataset --- the MRNet dataset which consists of around $30.000$ images of knees~\cite{dataset_knee}. To generate the matrix $W$ we again transform each training image into the DB4 wavelet basis and apply a threshold of about $0.006$ to get distribution of non-zero coefficients (top right). This time the resulting weights are non-symmetrical and hence plugging them into Formula~\eqref{formula} results in a non-symmetrical subsampling density, thereby \textit{adapting} to the underlying structure of the signals. This makes the difference between the adapted subsampling distribution and the polynomial subsampling strategy more pronounced, which will also result in greater differences in the PSNR. Sampling $10\%$ of measurements from the adapted and polynomial densities (bottom left and middle), we get by again applying the Nesta algorithm to~\eqref{BP} that our adapted subsampling scheme outperforms the heuristically inspired polynomial subsampling strategy --- resulting in a PSNR value of 27.9 compared to 26.8.\\
\noindent This difference in performance gets even more pronounced in the next experiment (Figure~\ref{fig:flip}), where we use the same setup (and dataset) as in the first experiment, but \textbf{flip} the sparse coefficients of each image (including the test image) by applying the transform $x \mapsto x^{f} \in \C^K$, $x^f_1 = x_K, x^f_2 = x_{K-1}, \cdots , x^f_K = x_1$ to the vectorised sparse coefficients. This is inspired by the so-called flip test~\cite{adhaporo17}. Obviously, the estimated distribution of the sparse supports is now flipped as well (top right) and plugging these weights $\poissonprob$ into Formula~\eqref{formula} yields a completely different sampling distribution. We again sample $10\%$ of measurements from the 2D k-space (bottom left and middle). This time, our adapted subsampling strategy easily outperforms the heuristic polynomial decay subsampling strategy --- resulting in a PSNR value of 22.7 compared to 12.0.\\
These experiments showed that our subsampling scheme really does adapt to the underlying distribution of sparse supports and outperforms heuristic subsampling schemes. 

\section{Special cases}\label{sec:special}
In this section we want to analyse our result for a few special cases of measurement matrices, sparsity basis and weights $\poissonprob$ which underline the generality of the above result. We further show how our result can be applied to recover state of the art theoretical results in CS theory.

\subsection{Coherent matrix}
A frequent example showing the necessity of some sort of knowledge of the structure in sparse signals is the special case $A_0 = \mathbb{I}$. Denote by $J := \{i : \poissonprob_i \neq 0\}$ the set of indices where the weights of our random support model are zero and set the columns of $A_{J^c}$ to zero. In this setting, Formula~\eqref{formula} leads to $\sampleprob_k = \frac{\delta_{k,J}}{\lvert J\rvert}$ and thus $m \gtrsim \lvert J \rvert \log^3(K/\varepsilon)$ which means that to ensure recovery with high probability, we have to sample all rows corresponding to positive weights $\poissonprob_{\ell}$, i.e. all those rows that correspond to entries of our sparse vector that have a non-zero probability of appearing in the support. This also includes the setting where $\poissonprob \in \{0,1\}$ recovering, up to logarithmic factors, results derived in~\cite{pierre15} for fixed sparse supports.

\subsection{Fourier matrix} Assume that $A_0 = \mathcal{F}$, i.e. $A_0$ is the $1$-D Fourier transform. This matrix is known to be incoherent ($\|a_k\|^2_{\infty} =  \frac{1}{K}$) and in the isolated measurement setting our result yields $ a_k D_{\poissonprob} a_k \transp =  \sum_{\ell} \lvert a_{k,\ell} \rvert \poissonprob_{\ell} \leq \|a_k\|_{\infty} \|\poissonprob\|_1= \frac{1}{K} \sum_{\ell} \poissonprob_{\ell} =  \frac{S}{K}$ for any weight vector $\poissonprob$ (recall that $\sum_{\ell} \poissonprob_{\ell} =S$). Plugging these observations back into our main Theorem yields that independent of the distribution $\poissonprob$, one should sample uniformly at random, i.e. $\sampleprob_k = \frac{1}{K}$. Corollary~\ref{lem} thus yields $m \gtrsim S \log^3(K)$ which (up to log factors) is in line with standard lower bounds on the number of measurements~\cite{candes06,donoho06}.

\subsection{Uniformly distributed sparse supports}
\begin{figure}
    \centering
    \includegraphics[width=0.66\linewidth]{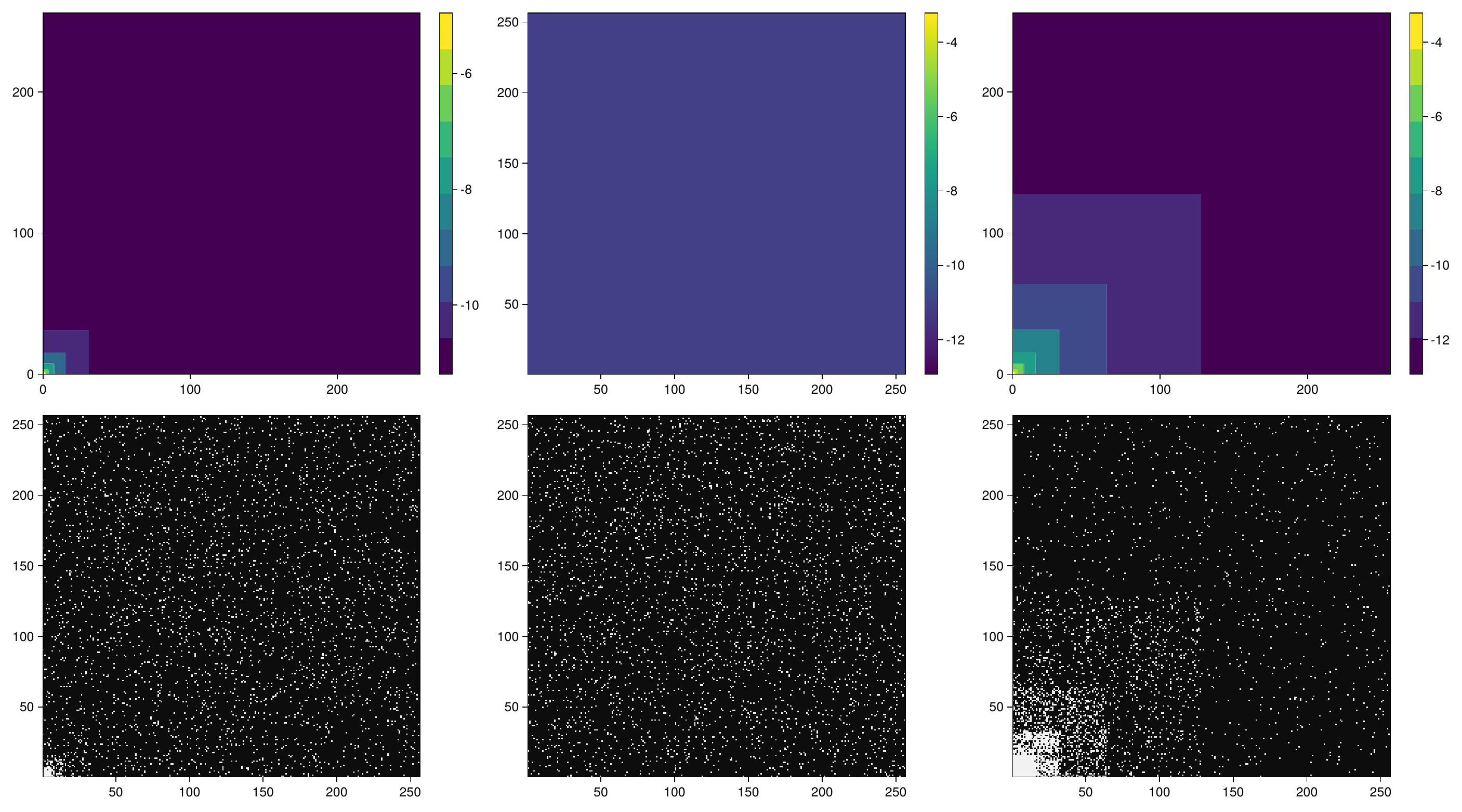}
    \caption{Subsampling densities (top row) and corresponding samples (bottom row) for the adapted variable density sampling scheme (left column), the uniform distribution (middle right) and the coherence based subsampling scheme (right row). The resulting average PSNR are: adapted - 133.5, uniform - 105.6 and coherence based- 62.3.}
    \label{fig:uniform}
\end{figure}

One possible distribution of our sparse supports is the uniform distribution, where $\poissonprob_{\ell} = S/K$. Plugging this into Formula~\eqref{formula} yields
\begin{align*}
   \sampleprob_k = \frac{\max\{S/K, \|a_k\|^2_{\infty}\}}{L}\label{uniform},
\end{align*}
where $L$ again is a normalising constant. This improves upon to coherence based subsampling strategies~\cite{ra10, weiss13, Vander11}, where $\sampleprob_k := \frac{\|a_k\|^2_{\infty}}{\sum_{\ell}\|a_{\ell}\|^2_{\infty}}$. Intuitively, since in the uniform case there is no structure in the sparse signals that can influence the subsampling strategy it is only natural that in this special case the optimal subsampling strategy depends only on the structure of the sensing matrix together with a lower bound $S/K$ to cover the whole space.\\
To see that the our result is indeed tight in the sense that both terms in the numerator of Formula~\ref{formula} are indeed necessary, we conduct a small experiment. We set $K = 2^{16}$, $S = \sqrt{K}/2$ and let $\Phi\transp$ be the 2D Hadamard transform and $\pdico$ be the 2D Haar wavelet transform. We then generate $100$ synthetic signals via our signal model~\ref{signal_model} with uniformly distributed sparse supports $\poissonprob_{\ell} = S/K$, coefficients $c = 1$ and random signs $\sigma = \pm 1$. We then compare the performance of three different subsampling strategies $\pi^1,\pi^2$ and $\pi^3$ defined as
\begin{align}
   \sampleprob^1_k = \frac{\max\{S/K, \|a_k\|^2_{\infty}\}}{L}, \quad \quad \quad  \sampleprob^2_k = \frac{m}{K},\quad \quad \quad\sampleprob^3_k =  \frac{\|a_k\|^2_{\infty}}{\sum_{\ell}\|a_{\ell}\|^2_{\infty}}.
\end{align}
We call them the adapted, uniform and coherence based subsampling strategies --- see also top row of Figure~\ref{fig:uniform}.
Sampling $5\%$ of measurements from each of these distributions and subsequently solving~\ref{BP} with the Nesta algorithm~\cite{Nesterov05smoothminimization, candes11_nesta} and averaging the PSNR over $10$ runs, each with $100$ fresh signals, shows that our adapted subsampling strategy outperforms both the uniform and the coherence bases subsampling strategy, indicating that both terms in~\eqref{uniform} are not only sufficient, but indeed necessary.
\subsection{Sparsity in levels}
A frequent assumption in modern compressed sensing theory is sparsity in levels~\cite{adhaporo17, pierre15,adcock20,adcock16,adcock19}. To apply our results to this framework we assume that $K = 2^{J+1}$ for some $J \in \N$ and set $A_0 = \mathcal{F}\pdico\transp$, where $\mathcal{F}$ is the $1$-D Fourier transform with rows indexed from $-K/2+1$ to $K/2$ and $\pdico$ is the $1$-D Haar wavelet transform. 
Denote by $\Omega$ the dyadic partition of the set $\{1, \cdots , K\}$ where $\Omega_0 := 0$, $\Omega_1 := 1$ and $\Omega_{j} := \{2^{j-1} +1, \cdots, 2^{j}\}$ for $j = 2, \cdots, J+1$. Further denote by $M$ the $J+1$ frequency bands of the discrete Fourier transform $\mathcal{F}$, i.e., $M_0 := \{0,1\}$ and $M_j := \{ -2^{j}+1, \cdots ,-2^{j-1} \} \cup \{ 2^{j-1}+1, \cdots ,2^{j} \}$ for $j = 1, \cdots, J$. Lemma 1 in~\cite{adcock16} states that for $\ell \in \Omega_{i}$ and $k \in M_{j}$
\begin{equation}
    \lvert a_{k,\ell} \rvert^2 \lesssim 2^{-j} 2^{\lvert j-i \rvert}.\label{four_wav_bound}
\end{equation}
We define the \textbf{average sparsity in level} $\ell$ as
\begin{align}
    S_{\ell} := \| \poissonprob_{\Omega_{\ell}} \|_1
\end{align}
For simplicity we assume $S_{\ell} > 1$ for all $1 \leq \ell \leq J$. Plugging this into~\eqref{formula} yields for $k \in M_j$
\begin{equation}
    a_k D_{\poissonprob} a_k \transp \lesssim 2^{-j} S_{j} + 2^{-j} \sum_{p \neq j} 2^{\lvert j - p \rvert}S_p,
\end{equation}
and thus by using $\sampleprob$ as defined in~\eqref{formula} our main result yields the sufficient condition
\begin{align}
    m \gtrsim \left(\sum_{j} S_{j} + \sum_{p \neq j} 2^{ \lvert j - p \rvert}S_p \right)\log^3(K/\varepsilon),
\end{align}
in line with results in~\cite{adcock20}. 
\subsection{Blocks of measurements}\label{sec:blocks}
Even though the above sampling strategies yield very good reconstruction results, probing measurements independently at random is infeasible --- or at least impractical --- in most real applications, see~\cite{pierre15} and references therein. Luckily, our results easily extend to the case of blocks of measurements $B_k$.
\subsubsection{Sensing vertical (or horizontal) lines in 2D}
We will again follow the notation in~\cite{pierre15,adcock20} very closely to facilitate easier comparison. Assume again that $K = 2^{J+1}$ for some odd $J \in \N$. Let $\dico \in \C^{\sqrt{K}\times \sqrt{K}}$ be a unitary matrix (for example the discrete $1$D Fourier-Haar transform matrix) and assume that our set of possible measurements is given by
\begin{equation}
    A_0 = \dico \otimes\dico \in \C^{K \times K},
\end{equation}
where $\otimes$ denotes the Kronecker product. With this notation, we define blocks of measurements which, in a 2D Fourier-Wavelet setting would correspond to vertical lines in frequency space. For this set
\begin{equation*}
    B_k := \dico_{k,:} \otimes \dico = \left( \dico_{k,1} \dico \; \lvert \dots    \rvert \; \dico_{k,\sqrt{K}} \dico   \right) \in \C^{\sqrt{K} \times K} \quad \text{for all} \quad 1 \leq k \leq \sqrt{K}.
\end{equation*}
The separable nature of this setup has the big advantage that the matrix $B_k \transp B_k$ has a very nice representation. Note that in our main result we have to control $\|B_k D_{\poissonprob} B_k\transp\| = \| \weightsp B_k \transp B_k \weightsp\|$. Using that $\dico$ is a unitary matrix we see
\begin{align}
    B_k \transp B_k = (\dico_{k,:} \otimes \dico) \transp (\dico_{k,:} \otimes \dico) = (\dico_{k,:} \transp \dico_{k,:} \otimes \dico \transp \dico) = (\dico_{k,:} \transp \dico_{k,:} \otimes \mathbbm{I}).
\end{align}
For our weight vector $\poissonprob \in \R^{K}$ we denote by $W \in \R^{\sqrt{K} \times \sqrt{K}}$ the matrix satisfying $\operatorname{vec}(W) = \poissonprob$. Multiplying $B_k \transp B_k = (\dico_{k,:} \transp \dico_{k,:} \otimes \mathbbm{I})$ from the left and right with the diagonal matrix $\weightsp$ and taking the operator norm yields
\begin{align*}
    \| \weightsp(\dico_{k,:} \transp \dico_{k,:} \otimes \mathbbm{I}) \weightsp\| = \| \weightsp \begin{pmatrix} \dico_{k,1}\transp \dico_{k,1} \mathbbm{I}& \dots  & \dico_{k,1}\transp \dico_{k,\sqrt{K}} \mathbbm{I}\\
    \vdots & \ddots & \vdots\\
    \dico_{k,\sqrt{K}}\transp \dico_{k,1} \mathbbm{I} & \dots  & \dico_{k,\sqrt{K}}\transp \dico_{k,\sqrt{K}} \mathbbm{I}  \end{pmatrix} \weightsp\|.
\end{align*}
Since reordering of columns and rows does not change the operator norm, we apply the permutation $R: \mathbb{K} \mapsto \operatorname{vec}(\operatorname{vec}^{-1}(\mathbb{K})\transp)$ to both the columns and rows of the above matrix and set $\poissonprob' := R(\poissonprob)$ to get
\begin{align*}
    \| \weightsp(\dico_{k,:} \transp \dico_{k,:} \otimes \mathbb{I}) \weightsp \| &= \left\| D_{\sqrt{\poissonprob'}}\begin{pmatrix} \dico_{k,:}\transp \dico_{k,:} & \dots  & 0\\
    \vdots & \ddots & \vdots\\
    0 & \dots  & \dico_{k,:}\transp \dico_{k,:} \end{pmatrix} D_{\sqrt{\poissonprob'}} \right\| \\&= \max_{1 \leq \ell \leq \sqrt{K}} \| \dico_{k,:} D_{W_{\ell,:}}^{1/2} \|^2_2 = \max_{1 \leq \ell \leq \sqrt{K}} \sum_{i = 1}^{\sqrt{K}} \lvert \dico_{k,i} \rvert^2 W_{\ell,i}.
\end{align*}
So we look for the row $v$ of the matrix $W$, such that $\| \dico_{k,:} D_{\sqrt{v}} \|^2_2$ is maximised. This encapsulates the relationship between the structure of the blocks of measurements and the structure of the sparse signals via their distribution. By the same argument as above we also see that
$
    \|B_k \transp B_k \|_{\infty,1} = \|\dico_k\|^2_{\infty}.
$
Plugging this into our formula for blocks~\eqref{formula_blocks} yields
\begin{align}
    \sampleprob_k := \frac{\max \left\{ \max_{1 \leq \ell \leq \sqrt{K}} \right\{ \sum_{i = 1}^{\sqrt{K}} \lvert\dico_{k,i}\rvert^2 W_{\ell,i} \left\}, \| \dico_k \|^2_{\infty} \right\}}{L},\label{sparsity_levels_blocks}
\end{align}
where $L$ is the normalisation constant. If instead of vertical lines one would take horizontal lines
$
    B_k := \dico \otimes \dico_{k,:},
$
we would get
\begin{align*}
    B_k \transp B_k = \begin{pmatrix} \dico_{k,:}\transp \dico_{k,:} & \dots  & 0\\
    \vdots & \ddots & \vdots\\
    0 & \dots  & \dico_{k,:}\transp \dico_{k,:} \end{pmatrix},
\end{align*}
without any reordering. Hence in this case
\begin{align*}
    \|\weightsp B_k \transp B_k \weightsp\| = \max_{1 \leq \ell \leq \sqrt{K}} \sum_{i = 1}^{\sqrt{K}} \lvert \dico_{k,i} \rvert^2 W_{i,\ell},
\end{align*}
which amounts to taking the maximum over all columns of the matrix $W$. Plugging this back into our formula for blocks~\eqref{formula_blocks} yields
\begin{align}
    \sampleprob_k := \frac{\max \left\{  \max_{1 \leq \ell \leq \sqrt{K}} \left\{ \sum_{i = 1}^{\sqrt{K}} \lvert \dico_{k,i} \rvert^2 W_{i,\ell} \right\},\| \dico_k \|^2_{\infty}   \right\}}{L},
\end{align}
where $L$ is again the normalisation constant.
\subsubsection{Vertical Fourier-Haar lines}
We now apply the above analysis to the special case where $\dico = \mathcal{F} \pdico\transp$ is the 1D Fourier-Haar transform. This yields that $A_0$ is the separable 2D Fourier-Haar transform\footnote{In all other experiments we use non-separable 2D wavelet transforms.}. Let $\poissonprob \in \R^K$ again be our weight vector and define the matrix $W \in \R^{\sqrt{K} \times \sqrt{K}}$ such that $\operatorname{vec}(W) = \poissonprob$. We again denote by $M_{\ell}$ the frequency bands of the one dimensional Fourier transform and by $\Omega_{\ell}$ the dyadic partition (see previous subsection). In the 2D setting we define the \textbf{average maximal sparsity in level} $\ell$ as
\begin{align}
    S_{\ell} := \max_k \| W_{k,\Omega_{\ell}} \|_1.
\end{align}
This is equivalent to the 1D case up to taking the maximum over all rows of the matrix $W$. Using~\eqref{four_wav_bound} and assuming that $S_{\ell} >1$ for all $1 \leq \ell \leq J$, the above analysis yields for $k \in M_{j}$
\begin{align}
    \| B_k \transp D_{\poissonprob} B_k \| & =  \max_{1 \leq \ell \leq \sqrt{K}} \sum_{i = 1}^{\sqrt{K}} \lvert \dico_{k,i} \rvert^2 W_{\ell,i} \leq  \sum_{i = 1}^{\sqrt{K}} \max_{1 \leq \ell \leq \sqrt{K}} \lvert \dico_{k,i}\rvert^2 W_{\ell,i} \label{bad_inequality}\\
    & \lesssim 2^{-j} S_{j} + 2^{-j} \sum_{p \neq j} 2^{\lvert j - p \rvert}S_p,
\end{align}
and thus by using $\sampleprob$ as defined in~\eqref{sparsity_levels_blocks} our main result yields the sufficient condition
\begin{align}
    m \gtrsim \left(\sum_{j} S_{j} + \sum_{p \neq j} 2^{\lvert j - p \rvert}S_p \right)\log^3(K/\varepsilon),
\end{align}
in line with results in~\cite{adcock20}. Note that the first inequality in~\eqref{bad_inequality} is rather crude and potentially loses a lot of information about the relationship between the matrix $W$ and the structure of the 2D Fourier-Haar matrix $A_0 = \mathcal{F} \pdico\transp$. This is why in our experiments we will stick with the quantity $\| B_k \transp D_{\poissonprob} B_k \| =  \max_{1 \leq \ell \leq \sqrt{K}} \sum_{i = 1}^{\sqrt{K}} \lvert \dico_{k,i} \rvert^2 W_{\ell,i}$.
\subsubsection{Numerical experiments of blocks of measurements Fourier - DB4}
\begin{figure}[ht]
    \centering
    \includegraphics[width=0.66\linewidth,height = 0.57\linewidth]{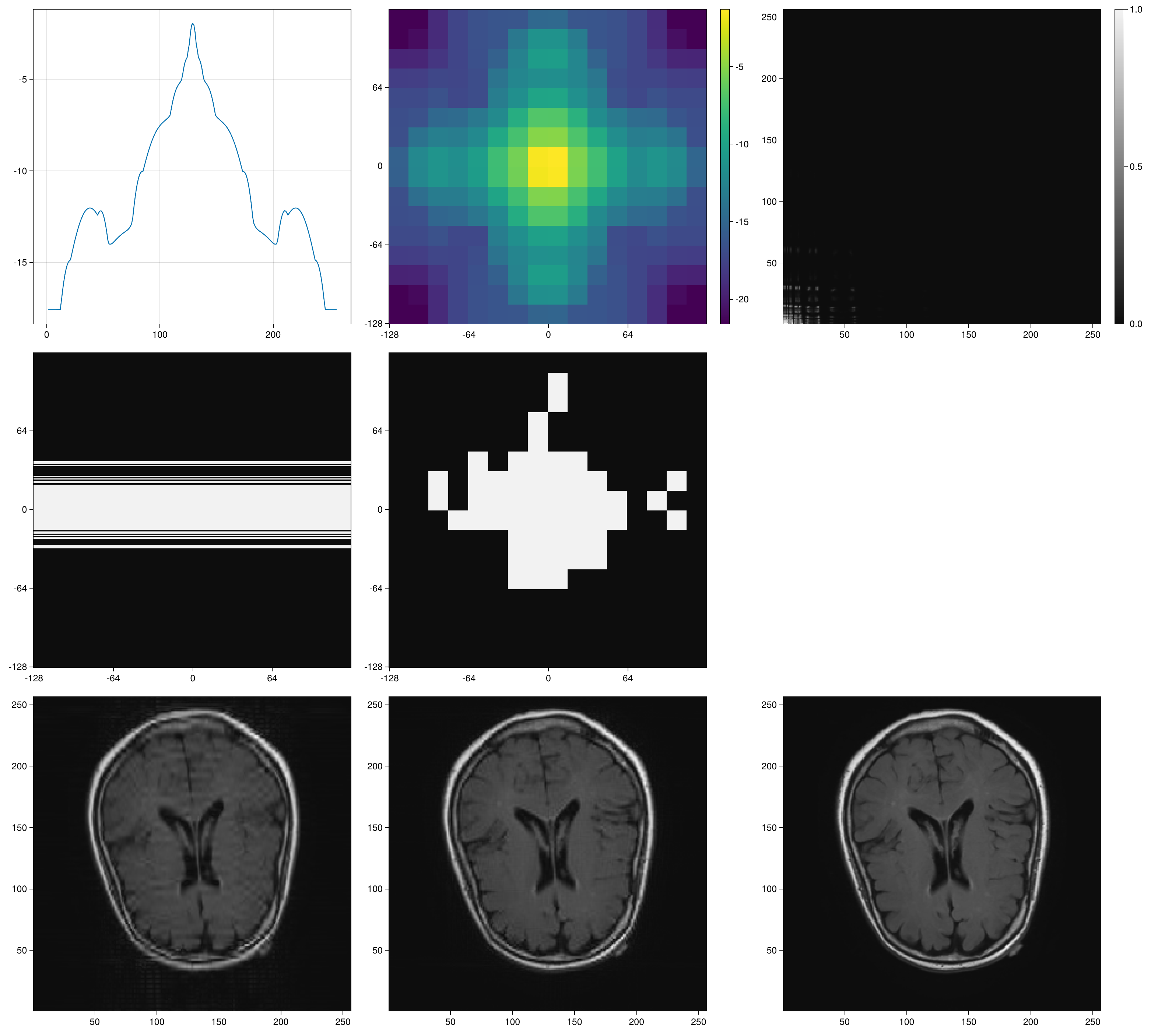}
    \caption{Adapted variable density sampling schemes with vertical lines (left column) and squares (middle column). Matrix $W$ of sparse support distribution in the separable 2D DB4 wavelet basis (top right), test image (bottom right) and reconstructions (bottom left and middle). The resulting PSNR values are: Lines - 29.9 and Squares - 33.9.}
    \label{fig:lines}
\end{figure}

In this subsection we will use blocks of measurements in numerical experiments--- Figure~\ref{fig:lines}. We conduct two experiments, first by measuring along horizontal lines in the 2D k-space (left column) and then by measuring square blocks of size $16 \times 16$ in the 2D k-space (middle column). We again use the Brain dataset with a threshold of around $0.023$ to generate a estimate of the matrix $W$ in the \textit{separable} 2D DB4 wavelet basis (top right). Plugging these estimated weights into Formula~\eqref{formula_blocks} we get an adapted sampling distribution on the vertical lines (top left) and on the square blocks (top middle). Sampling 20$\%$ of measurements from the 2D k-space (middle row) we get good reconstruction of the reference image (bottom right) for both measurement techniques (bottom left and middle). This shows how our results also apply to the setting of blocks of measurements.
\section{Discussion}\label{sec:discussion}
The above results showed that the optimal variable density subsampling strategy in a compressed sensing setup should not only depend on the structure of the sensing and sparsity matrices, but also on the distribution of sparsity patterns of the signals to be measured. We derived lower bounds on the number of measurements to ensure recovery of the sparse signals with high probability and derived a simple formula for the optimal subsampling strategy. We showed that this distribution can be estimated from a training set and that the resulting adapted subsampling scheme provides state of the art performance in a range of situations. For future work it would be interesting to analyse different settings of blocks of measurements, where explicit lower bounds on the number of measurements can be derived.
\section*{Acknowledgements}
This work was supported by the Austrian Science Fund (FWF) under Grant no.~Y760. The author wants to thank Karin Schnass for proof-reading the manuscript and her insightful comments. 

\appendix

\section{Proof of Theorem~\ref{them:1}}\label{sec:proof}
Now we turn to proving Theorem~\ref{them:1}. Note that we have three sources of randomness: the signs $\sigma$, the set of random measurements $J$ and the random supports $I$. Strictly speaking, we are working on the product measure of the three, but in slight abuse of notation, we will write $\P_{\sigma}$, $\P_{J}$ and $\P_S$ to indicate the probability measure that we use for the corresponding concentration inequalities. The exact statement of Theorem~\ref{them:1} --- including constants --- reads as
\begin{theorem}
Assume that $m$ measurements $B_k$ are sampled according to probabilities $\sampleprob_k$ and the signals follow the model in~\ref{signal_model}, where the support $I \subseteq \mathbb{K}$ is chosen according to the rejective sampling model with probabilities $\poissonprob_1, \cdots, \poissonprob_K$ such that $\sum_{k =1}^K \poissonprob_k = S$ and $0 < \poissonprob_k \leq 1$. If
\begin{align}
    m &\geq \max_{k}  \frac{\| B_{k}\transp B_{k} \|_{\infty,1} }{\sampleprob_{k}} 384 \log^2(10^4 K/\varepsilon)\log(54 K^2 /\varepsilon), \quad \text{and} \nonumber\\
    m &\geq \max_{k} \frac{\| B_{k} D_{\poissonprob} B_{k}\transp \|_{2,2}}{\sampleprob_{k}}   10^4 e^2 \log^2(10^4 K/ \varepsilon) ,
\end{align}
then \eqref{BP} recovers the sparse signal with probability $1- \varepsilon$.
\end{theorem}
Before we can state the proof, we need to establish $5$ concentration inequalities. To that end, we define
\begin{align*}
    \Lambda &:= \max_{k} \frac{\| B_{k} D_{\poissonprob} B_{k}\transp \|_{2,2}}{\sampleprob_{k}m}\quad \text{and} \quad
    \kappa := \max_{k}  \frac{\| B_{k}\transp B_{k} \|_{\infty,1} }{\sampleprob_{k}m}.
\end{align*}
Further let $H :=  A \transp A  -  \diag(A\transp A)$. We begin by bounding the biggest entry of $H$. For that we can apply the standard Bernsteins's inequality to each entry together with a unit bound to get
\begin{lemma}\label{lem_j2}
Let $A$ depend on the draw of the $j_{\ell}$. Then for all $t \geq 0$, we have
\begin{align}
    \P_{J}\left(\|A \transp A - \mathbbm{I} \|_{\infty,1} \geq t \right) \leq 2 K^2 \exp{\left( -\frac{t^2/ 2}{\kappa(1+t)/3} \right) }.
\end{align}  
\end{lemma}
\begin{proof} We fix indices $i$ and $j$ and bound the $i,j$-th entry of the matrix $A \transp A - \mathbbm{I}$. Recall that by definition of $A$
\begin{align}
       A \transp A  -  \mathbbm{I} = \sum_{k = 1}^m \frac{1}{m}\left(\frac{ B_{j_k} \transp B_{j_k} }{\sampleprob_{j_k}} - \mathbbm{I} \right).
\end{align}
Focusing only on the $i,j$-th entry of this matrix, we can write this entry as $\sum_{k = 1}^m x_k$, where $x_k := \frac{1}{m} e_j\transp \left(\frac{ B_{j_k} \transp B_{j_k} }{\sampleprob_{j_k}} - \mathbbm{I} \right) e_i$. By definition $\E[x_k] = 0$ and we can bound each $x_k \leq \kappa$ for all $k$. To bound the variance, note 
\begin{align}
         \E [ x_k ^2] &= \E \left[ \left(\frac{ e_j\transp B_{j_k}\transp B_{j_k}e_i}{\sampleprob_{j_k} m }\right)^2\right] - \frac{1}{m^2} 
   e_j \transp \; \mathbbm{I} \;  e_i  \leq \kappa \; \E \left[ \frac{ e_i \transp B_{j_k}\transp B_{j_k} e_i }{\sampleprob_{j_k}m}\right]\leq \kappa \frac{1}{m},
 \end{align}
which leads to $\sigma^2 =  \sum_{k =1 }^m \E[ x_k^2] \leq \kappa$. An application of the Bernstein inequality together with a union bound over all pairs $i,j$ yields the result.
\end{proof}
In the next step we want to bound the largest $\ell_2$-norm of the matrix $H \weightsp$, i.e. we want to bound $\|H \weightsp \|_{\infty,2}$. To that end we are going to apply the vector Bernstein inequality~\cite{MINSKER2017111} together with a union bound. Concretely we have
\begin{lemma}\label{lem_j1}
Let $I$ be a fixed support of cardinality $S$ and let $A$ depend on the draw of the $j_{\ell}$. Then for all $t>0$, we have
\begin{align*}
    \P_J \left(\max_{i}\|  A_i \transp A_{/i} \weightsp  \|_{2} \geq t \right) \leq 28 K \exp{\left( -\frac{t^2/ 2}{\Lambda + \sqrt{\Lambda \kappa} \cdot t/3} \right)}.
\end{align*}
\end{lemma}
\begin{proof}
For a fixed $i$ we set $D_{/i}$ the diagonal matrix that is one except for the $i$-th entry, where it is zero. For ease of notation we are going to look at the transposed vector $\weightsp A_{/i} \transp A_i$ which we can write as
\begin{equation*}
    \| \weightsp D_{/i} A \transp A e_i \|_2 = \|  \sum_{k = 1}^m  \frac{ \weightsp D_{/i}  B_{j_k} \transp B_{j_k} e_i}{\sampleprob_{j_k} m }  \|_2 =  \| \sum_{k=1}^m X_k \|_2,
\end{equation*}
where $$X_k := \frac{\weightsp D_{/i} B_{j_k} \transp B_{j_k} e_i}{\sampleprob_{j_k} m }.$$By definition we have $\E[X_k] =  \frac{1}{m} \weightsp D_{/i} \sum_{\ell =1}^M B_{\ell} B_{\ell} \transp e_i = \frac{1}{m} \weightsp D_{/i} e_i =  0$. Further
\begin{align}
    \max_k \|X_k\|_2 &= \max_k \left\| \frac{\weightsp B_{k} \transp B_{k}e_i }{\sampleprob_{k} m} \right\|_2 \nonumber \leq \sqrt{\Lambda \kappa}
\end{align}
To bound the variance, note that
\begin{align*}
    \E[\|X_k\|_2^2] &= \E \left[ \left\| \frac{\weightsp B_{j_k} \transp B_{j_k} e_i }{\sampleprob_{j_k} m }\right\|_2^2 \right]  \leq \Lambda \E \left[ \left\| \frac{B_{j_k} e_i}{\sqrt{\sampleprob_{j_k} m }} \right\|_2^2 \right]  = \Lambda \|e_i\|_2^2 \frac{1}{m} =  \frac{\Lambda}{m}.
\end{align*}
This leads to $\sigma^2 = \sum_{k = 1}^m \E [\|X_k \|_2^2] \leq \Lambda$. An application of the vector Bernstein inequality~\cite{MINSKER2017111} together with an union bound finishes the proof.
\end{proof}
Next we show how to bound the operator norm of the matrix $\weightsp H \weightsp$. The following result is similar to standard results in CS theory~{\cite[Lemma 2.1]{candes11}} and~{\cite[Lemma C.1]{pierre15}}. The Matrix Bernstein inequality~\cite{tr12} yields
\begin{lemma}\label{lem_j3}
Let $A$ depend on the draw of the $j_{\ell}$. Then for all $t \geq 0$, we have
\begin{align*}
    \P_{J}\left(\| \weightsp A \transp A  \weightsp - \diag(A \transp A ) D_p \|_{2,2} \geq t \right) \leq 2 K \exp{\left( -\frac{t^2/ 2}{ 4 \Lambda(1 + t)/3} \right)}.
\end{align*}
\end{lemma}
\begin{proof} We begin by noting that we can write
\begin{align}
    \weightsp H \weightsp &=  \weightsp A \transp A \weightsp -  \diag(A \transp A) D_p \nonumber\\
    &= \sum_{k = 1}^m \left(\frac{\weightsp B_{j_k} \transp B_{j_k}\weightsp }{\sampleprob_{j_k} m } - \diag\left( \frac{ B_{j_k} \transp B_{j_k}  }{\sampleprob_{j_k} m } \right) D_p \right)  = \sum_{k=1}^m X_k,
\end{align}
where $X_k := \left(  \frac{\weightsp B_{j_k} \transp B_{j_k} \weightsp }{\sampleprob_{j_k} m }-\diag\left( \frac{ B_{j_k} \transp B_{j_k}  }{\sampleprob_{j_k} m } \right)  D_p\right)$. By definition of the $j_k$ we have $\E[X_k] = 0$. Further
\begin{equation*}
    \| X_k \|_{2,2} \leq 2 \max_k \frac{\| \weightsp B_{k}\transp B_{k} \weightsp \|_{2,2}}{\sampleprob_{k} m } =  2 \Lambda.
\end{equation*}
Using that for two discrete symmetric random matrices $A$ and $B$ over the same probability space that we have~\cite{ru22diss}
\begin{align}
\| \E [ \big(A &+ B\big) \big(A + B\big)\transp]\| \leq \left(\| \E [ A^2 ]\|^{\frac{1}{2}} +  \| \E [ B^2]\|^{\frac{1}{2}}\right)^2.  
\end{align}
Using this to bound the variance we get
\begin{align*}
   \| \E [ X_k ^2] \|  & \leq 4 \| \E \left[ \left(\frac{ \weightsp B_{j_k}\transp B_{j_k}\weightsp}{\sampleprob_{j_k} m }\right)^2\right] \| \leq 4 \Lambda \frac{1}{m},
\end{align*}
which leads to $\sigma^2 = \| \sum_{k =1 }^m \E[ X_k^2] \|_{2,2} \leq  4 \Lambda$. An application of the Matrix Bernstein inequality yields the result.
\end{proof}
We further need the following Hoeffding-like tail bound for sums of centered complex random variables --- see~(\cite{foucart13} Corollary 7.21 and Corollary 8.10).
\begin{lemma}\label{bern}
    Let $M \in \C^{K \times S}$ be a matrix and $x \in \R^{S}$ such that $\signop(x_i) \in \R^{S}$ is an independent Rademacher sequence. Then, for all $t \geq 0$
\[
\P_{\sigma}\left(\|M x\|_{\infty } \geq t\right) \leq 2K \exp{\left( -\frac{t^2}{2 \| M \|_{\infty,2}^2\| x\|_{\infty}^2} \right)}.
\]
\end{lemma}
The following concentration inequality can be found in~\cite{rusc21} and follows from a decoupling argument followed by an application of the standard Chernoff inequality and a union bound.
\begin{lemma}[{\cite[Lemma~3.4]{rusc21}}]\label{CS:lem:4}
Let $H \in \C^{K \times K}$ be some matrix. Assume $I \subseteq \mathbb{K}$ is chosen according to the rejective sampling model with probabilities $\poissonprob_1, \dots , \poissonprob_K$ such that $\sum_{i = 1}^K \poissonprob_i = S$. Further let $\poissonprob$ denote the corresponding weight vector. Then, for all $v > 0$
\begin{equation*}
    \P_S \left( \| H_I \|_{\infty,2} \geq v \right) \leq 2 K \left( e \frac{\| H \weightsp \|_{\infty,2}^2}{v^2}  \right) ^{\frac{v^2}{\|H\|_{\infty,1}^2}}.
\end{equation*}
\end{lemma}
The key ingredient to prove Theorem~\ref{them:1} is the following concentration inequality for the operator norm of random submatrices with non-uniformly distributed supports which can be found in~\cite{rusc21}\footnote{The result in the cited paper is stated only for real matrices, but a careful analysis of the proof shows that this result also holds for complex matrices.}. This is what allows us to go one step further than existing results in analysing the underlying relationship between the sensing matrix and the distribution of sparse supports.
\begin{lemma}[{\cite[Theorem~3.1]{rusc21}}]\label{CS:lem:5}
Let $H \in \C^{K \times K}$ be a matrix with zero diagonal. Assume that the support $I \subseteq\mathbb{K}$ is chosen according to the rejective sampling model with probabilities $\poissonprob_1, \dots , \poissonprob_K$ such that $\sum_{i = 1}^K \poissonprob_i = S$. Further let $\poissonprob$ denote the corresponding weight vector. If $t \geq 2 e^2 \|\weightsp H \weightsp\|_{2,2}$, then
\begin{equation}
    \P(\|H_{I,I}\|_{2,2} \geq t ) \leq 216 K \exp{ \left( - \min \left \{ \frac{t^2}{4 e^2 \| H \weightsp\|^2_{\infty,2} }, \frac{t}{2 \| H \|_{\infty,1}}     \right\} \right)}.
\end{equation}
\end{lemma}
With all of these concentration inequalities in place we are finally able to prove Theorem~\ref{them:1}.
\begin{proof}
From~\cite{troppl1, Fuchs2004} we know that if $\|A_{I^c} \transp A_I (A_I\transp A_I)^{-1}  \sigma_I\|_{\infty} < 1$, then $x$ is the unique solution of the $\ell_1$-minimisation problem~\eqref{BP}. 
Set $M := A_{I^c} \transp A_I (A_I \transp A_I)^{-1}$ and assume that $ \| A_I \transp A_I - \mathbb{I} \| \leq 1/2$. Then
\begin{align*}
\| M & \|_{\infty,2}  =  \| A_{I^c} \transp A_I (A_I\transp A_I)^{-1}  \|_{\infty,2}  \leq \| A_{I^c} \transp A_I \|_{\infty,2} \| (A_I\transp A_I)^{-1}\|_{2,2}
\leq 2 \| A_{I^c} \transp A_I \|_{\infty,2} .
\end{align*}
Noting that $\| A_{I^c} \transp A_I \|_{\infty,2} = \max_{i \in I^c}\| A_{I} \transp A_i \|_{2}$ we have
\begin{align*}
    \P \left( \| M \sigma \|_{\infty}   \geq 1  \right) & \leq \P_{\sigma} \left( \| M \sigma \|_{\infty} \geq 1 \; \middle\lvert \;  \|M\|_{\infty,2}  \leq 2 \gamma \right) \\
    & \quad + \P \left( \| A_I \transp A_I - \mathbb{I} \|_{2,2} \geq 1/2\right) + \P \left(\max_{i \in I^c}\| A_{I} \transp A_i \|_{2} \geq \gamma \right).
\end{align*}
Setting $\gamma^2 =  \frac{1}{8 \log(6 K/ \varepsilon)}$ and applying Lemma~\ref{bern} to $M\sigma$ yields that the first term on the right hand side is bound by $\varepsilon/3$. We denote by $J := (j_{1},\dots,j_{m} )$ the $m$-dimensional vector that selects which blocks of measurements are taken. Setting $H := A \transp A - \diag (A \transp A)$ we see that $H$ depends on the random sequence $J$. With that in mind we define 
\begin{align}
    \mathcal{J} := \left\{ J  \; \middle\lvert \; \|H\|_{\infty,1} \leq v_1, \| H \weightsp\|^2_{\infty,2} \leq v_2    , \|\weightsp H \weightsp\|_{2,2} \leq v_3 \right \}
\end{align}
to be the collection of sequences $J$ that satisfy the above conditions. Further
\begin{align}
    &\P \left(\| A_I \transp A_I - \mathbb{I} \|_{2,2} \geq 1/2\right) + \P \left(\max_{i \in I^c}\| A_{I} \transp A_i \|_{2} \geq \gamma \right) \nonumber \\
    &\quad \quad \quad \leq \P_{S} \left(\| A_I \transp A_I - \diag(A_I \transp A_I) \|_{2,2}\geq 1/4 \; \middle\lvert \; \mathcal{J} \right) 
    + \P_J( \mathcal{J} ) \nonumber \\
    &\quad \quad \quad \quad \quad \quad + \P \Big(\| \diag(A_I \transp A_I) - \mathbb{I}\|_{2,2} \geq 1/4  \Big) \nonumber \\
    &\quad \quad \quad \quad \quad \quad + \P_S \left(\max_{i \in I^c}\| A_{I} \transp A_i \|_{2} \geq \gamma \; \middle\lvert \; \mathcal{J}\right) + \P_J \left( \mathcal{J} \right)\label{eq:02} 
\end{align}
By Lemma~\ref{CS:lem:4} and Lemma~\ref{CS:lem:5} we can bound the two conditional probabilities via (for $v_3$ defined as below)
\begin{align}
     &\P_{S} \left(\| A_I \transp A_I - \diag(A_I \transp A_I ) \|_{2,2} \geq 1/4 \; \middle\lvert \; \mathcal{J} \right)    + \P_S \left(\max_{i \in I^c}\| A_{I} \transp A_i \|_{2} \geq \gamma \; \middle\lvert  \; \mathcal{J} \right) \nonumber \\
    & \quad \leq  216 K \exp{ \left( - \min \left \{ \frac{(1/4)^2}{4 e^2 v_2^2 }, \frac{1/4}{2 v_1}     \right\} \right)} + 2 K \left( e \frac{v_2^2}{\gamma^2}  \right) ^{\frac{\gamma^2}{v_1^2}}.\label{eq:01}
\end{align}
Setting
\begin{align}
    v_1 & : = \frac{1}{8 \log(216*6 K /\varepsilon)}   \nonumber \\
    v_2 & : = \frac{1}{8 e \log^{1/2}(216*6 K /\varepsilon)}\nonumber \\
    v_3 & : = \frac{1}{8 e^2}
\end{align}
we get that~\eqref{eq:01} is smaller than $\varepsilon/3$. Plugging this into~\eqref{eq:02} we see that to finish the proof we have to show that $\P \Big(\| \diag(A_I \transp A_I) - \mathbb{I}\| \geq 1/4  \Big)+ 2 \P (\mathcal{J}) \leq \varepsilon/3$. Using Lemmas~\ref{lem_j1}, \ref{lem_j2} and~\ref{lem_j3} to bound the three terms in $\mathcal{J}$ we get
\begin{align}
    \P \Big(\| \diag(A_I \transp A_I) - & \mathbb{I}\|_{2,2} \geq 1/4  \Big) + 2 \P_J(\mathcal{J})  \leq 6 K^2 \exp{\left( -\frac{v_1^2/ 2}{\kappa(1+v_1)/3} \right) } 
     \nonumber \\
    & + 56 K \exp{\left( -\frac{v_2^2/ 2}{\Lambda + \sqrt{\Lambda \kappa} \cdot v_2/3} \right)} + 4 K \exp{\left( -\frac{v_3^2/ 2}{4 \Lambda(1 + v_3)/3} \right)} \nonumber \\
    & \leq 
    6 K^2 \exp{\left( -\frac{v_1^2}{4 \kappa} \right) }  + 56 K \exp{\left( -\frac{v_2^2}{4 \Lambda } \right)} 
      + 4 K \exp{\left( -\frac{v_3^2}{2.72 \Lambda} \right)}.
\end{align}
By the assumptions on $\Lambda$ and $\kappa$ we indeed have $3 \P_J(\mathcal{J}) \leq \varepsilon/3$ which finishes the proof.
\begin{remark}
The proof of our main result relies heavily on the random signs of our signals. One could remove this assumption by instead employing the so-called "golfing scheme" proposed in~\cite{gr11}. Following the argument in~\cite{candes11} one should be able to derive similar results in the case of deterministic sign patterns. Since this would not have any impact on the optimal sampling distribution we opted for the shorter proof presented here.
\end{remark}
\end{proof}


%

\bibliography{karinbibtex.bib}

\begin{thebibliography}{30}
\providecommand{\natexlab}[1]{#1}
\providecommand{\url}[1]{\texttt{#1}}
\expandafter\ifx\csname urlstyle\endcsname\relax
  \providecommand{\doi}[1]{doi: #1}\else
  \providecommand{\doi}{doi: \begingroup \urlstyle{rm}\Url}\fi

\bibitem[Adcock et~al.(2016)Adcock, Hansen, and Roman]{adcock16}
B.~Adcock, A.C. Hansen, and B.~Roman.
\newblock A note on compressed sensing of structured sparse wavelet
  coefficients from subsampled fourier measurements.
\newblock \emph{IEEE Signal Processing Letters}, 23\penalty0 (5):\penalty0
  732--736, 2016.

\bibitem[Adcock et~al.(2017)Adcock, Hansen, Poon, and Roman]{adhaporo17}
B.~Adcock, A.C. Hansen, C.~Poon, and B.~Roman.
\newblock Breaking the coherence barrier: A new theory for compressed sensing.
\newblock \emph{Forum of Mathematics, Sigma}, 5:\penalty0 e4, 2017.

\bibitem[Adcock et~al.(2020)Adcock, Boyer, and Brugiapaglia]{adcock20}
B.~Adcock, C.~Boyer, and S.~Brugiapaglia.
\newblock On oracle-type local recovery guarantees in compressed sensing.
\newblock \emph{Information and Inference: A Journal of the IMA}, 10\penalty0
  (1):\penalty0 1--49, 2020.

\bibitem[Becker et~al.(2011)Becker, Bobin, and {C}and{\`e}s]{candes11_nesta}
S.~Becker, J.~Bobin, and {E}. {C}and{\`e}s.
\newblock Nesta: A fast and accurate first-order method for sparse recovery.
\newblock \emph{SIAM Journal on Imaging Sciences}, 4\penalty0 (1):\penalty0
  1--39, 2011.

\bibitem[Bigot et~al.(2013)Bigot, Boyer, and Weiss]{bibocl14}
J.~Bigot, C.~Boyer, and P.~Weiss.
\newblock An analysis of block sampling strategies in compressed sensing.
\newblock \emph{IEEE Transactions on Information Theory}, 62, 05 2013.

\bibitem[Boyer et~al.(2019)Boyer, Weiss, and Bigot]{pierre15}
C.~Boyer, P.~Weiss, and J.~Bigot.
\newblock Compressed sensing with structured sparsity and structured
  acquisition.
\newblock \emph{Applied and Computational Harmonic Analysis}, 46\penalty0
  (2):\penalty0 312 -- 350, 2019.

\bibitem[Buda(2019)]{dataset_brains}
M.~Buda.
\newblock Brain {MRI} segmentation.
\newblock
  \url{https://www.kaggle.com/datasets/mateuszbuda/lgg-mri-segmentation}, 2019.
\newblock Accessed: 2022-06-27.

\bibitem[{C}and{\`e}s et~al.(2006){C}and{\`e}s, Romberg, and Tao]{candes06}
{E}. {C}and{\`e}s, J.~Romberg, and T.~Tao.
\newblock Robust uncertainty principles: exact signal reconstruction from
  highly incomplete frequency information.
\newblock \emph{IEEE Transactions on Information Theory}, 52\penalty0
  (2):\penalty0 489--509, 2006.

\bibitem[Candes and {Plan}(2011)]{candes11}
E.~J. Candes and Y.~{Plan}.
\newblock A probabilistic and ripless theory of compressed sensing.
\newblock \emph{IEEE Transactions on Information Theory}, 57\penalty0
  (11):\penalty0 7235--7254, 2011.

\bibitem[Chauffert et~al.(2013)Chauffert, Ciuciu, Kahn, and Weiss]{weiss13_2}
N.~Chauffert, P.~Ciuciu, J.~Kahn, and P.~Weiss.
\newblock Variable density sampling with continuous trajectories.
\newblock \emph{SIAM Journal on Imaging Sciences}, 7, 11 2013.

\bibitem[{Chauffert} et~al.(2013){Chauffert}, {Ciuciu}, and {Weiss}]{weiss13}
N.~{Chauffert}, P.~{Ciuciu}, and P.~{Weiss}.
\newblock Variable density compressed sensing in {M}{R}{I}. {T}heoretical vs
  heuristic sampling strategies.
\newblock In \emph{2013 IEEE 10th International Symposium on Biomedical
  Imaging}, pages 298--301, 2013.

\bibitem[Chun and Adcock(2017)]{adcock17}
I.Y. Chun and B.~Adcock.
\newblock Compressed sensing and parallel acquisition.
\newblock \emph{IEEE Transactions on Information Theory}, 63\penalty0
  (8):\penalty0 4860--4882, 2017.

\bibitem[Donoho(2006)]{donoho06}
D.L. Donoho.
\newblock Compressed sensing.
\newblock \emph{IEEE Transactions on Information Theory}, 52\penalty0
  (4):\penalty0 1289--1306, 2006.

\bibitem[Donoho et~al.(2006)Donoho, Elad, and Temlyakov]{doelte06}
D.L. Donoho, M.~Elad, and V.N. Temlyakov.
\newblock Stable recovery of sparse overcomplete representations in the
  presence of noise.
\newblock \emph{{IEEE} {T}ransactions on {I}nformation {T}heory}, 52\penalty0
  (1):\penalty0 6--18, January 2006.

\bibitem[Foucart and Rauhut(2013)]{foucart13}
S.~Foucart and H.~Rauhut.
\newblock \emph{A Mathematical Introduction to Compressive Sensing}.
\newblock Applied and Numerical Harmonic Analysis. Birkhäuser, 2013.

\bibitem[Fuchs(2004)]{Fuchs2004}
J.~Fuchs.
\newblock On sparse representations in arbitrary redundant bases.
\newblock \emph{IEEE Transactions on Information Theory}, 50:\penalty0
  1341--1344, 2004.

\bibitem[Gross(2011)]{gr11}
D.~Gross.
\newblock Recovering low-rank matrices from few coefficients in any basis.
\newblock \emph{{IEEE} {T}ransactions on {I}nformation {T}heory}, 57\penalty0
  (3):\penalty0 1548--1566, 2011.

\bibitem[Hajek(1964)]{hajek1964}
J.~Hajek.
\newblock Asymptotic theory of rejective sampling with varying probabilities
  from a finite population.
\newblock \emph{Annals of Mathematical Statistics}, 35\penalty0 (4):\penalty0
  1491--1523, 1964.

\bibitem[HRauhut(2010)]{ra10}
H.~HRauhut.
\newblock \emph{Compressive Sensing and Structured Random Matrices}, pages
  1--92.
\newblock De Gruyter, Berlin, New York, 2010.

\bibitem[Krahmer and Ward(2012)]{krwa12}
F.~Krahmer and R.~Ward.
\newblock Stable and robust sampling strategies for compressive imaging.
\newblock \emph{arXiv:1210.2380}, 2012.

\bibitem[Li and Adcock(2019)]{adcock19}
C.~Li and B.~Adcock.
\newblock Compressed sensing with local structure: Uniform recovery guarantees
  for the sparsity in levels class.
\newblock \emph{Applied and Computational Harmonic Analysis}, 46\penalty0
  (3):\penalty0 453--477, 2019.

\bibitem[Minsker(2017)]{MINSKER2017111}
S.~Minsker.
\newblock On some extensions of {B}ernstein’s inequality for self-adjoint
  operators.
\newblock \emph{Statistics and Probability Letters}, 127:\penalty0 111--119,
  2017.

\bibitem[Nesterov(2005)]{Nesterov05smoothminimization}
Y.~Nesterov.
\newblock Smooth minimization of nonsmooth functions.
\newblock \emph{Math. Programming}, pages 127--152, 2005.

\bibitem[{Puy} et~al.(2011){Puy}, {Vandergheynst}, and {Wiaux}]{Vander11}
G.~{Puy}, P.~{Vandergheynst}, and Y.~{Wiaux}.
\newblock On variable density compressive sampling.
\newblock \emph{IEEE Signal Processing Letters}, 18\penalty0 (10):\penalty0
  595--598, 2011.

\bibitem[Ruetz(2022)]{ru22diss}
S.~Ruetz.
\newblock \emph{Compressed Sensing and Dictionary Learning with Non-Uniform
  Support Distribution}.
\newblock PhD thesis, University of Innsbruck, 2022.

\bibitem[Ruetz and Schnass(2021)]{rusc21}
S.~Ruetz and K.~Schnass.
\newblock Submatrices with non-uniformly selected random supports and insights
  into sparse approximation.
\newblock \emph{SIAM Journal on Matrix Analysis and Applications}, 42\penalty0
  (3):\penalty0 1268--1289, 2021.

\bibitem[Ruetz and Schnass(2022)]{ruetz2022nonasymptotic}
S.~Ruetz and K.~Schnass.
\newblock Non-asymptotic bounds for inclusion probabilities in rejective
  sampling, 2022.

\bibitem[{Standford {ML} group}(2019)]{dataset_knee}
{Standford {ML} group}.
\newblock {MRN}et {D}ataset.
\newblock \url{https://stanfordmlgroup.github.io/competitions/mrnet/}, 2019.
\newblock Accessed: 2022-06-27.

\bibitem[{Tropp}(2005)]{troppl1}
J.~{Tropp}.
\newblock Recovery of short, complex linear combinations via l1 minimization.
\newblock \emph{IEEE Transactions on Information Theory}, 51:\penalty0
  1568--1570, 2005.

\bibitem[{T}ropp(2012)]{tr12}
{J}. {T}ropp.
\newblock User-friendly tail bounds for sums of random matrices.
\newblock \emph{Foundations of Computational Mathematics}, 12\penalty0
  (4):\penalty0 389--434, 2012.

\end{thebibliography}
\end{document}